\documentclass[lenghtcheck,twocolumn,subeqn,amsmath,floats,tabularx,unsortedaddress,superscriptaddress,nofootinbib,prl,amssymb]{revtex4}
\usepackage{graphicx}
\usepackage{tabularx}  

\begin{document}
\hyphenation{mo-des}
\title{Nonequilibrium steady state fluctuations in actively cooled resonators}
\author{M. Bonaldi}
\email[Electronic mail: ]{bonaldi@science.unitn.it}
\affiliation{Istituto di Fotonica e Nanotecnologie, CNR-Fondazione Bruno Kessler, 38100 Povo, Trento, Italy}
\affiliation{INFN, Gruppo Collegato di Trento, Sezione di Padova, 38100 Povo, Trento, Italy}

\author{L. Conti}
\email[Electronic mail: ]{Livia.Conti@lnl.infn.it}
\affiliation{INFN, Sezione di Padova, Via Marzolo 8, 35131 Padova, Italy}

\author{P. De\;Gregorio}
\affiliation{Dip. di Matematica, Politecnico di Torino, Corso Duca degli Abruzzi 24, 10129 Torino, Italy}

\author{L. Rondoni}
\affiliation{Dip. di Matematica, Politecnico di Torino, Corso Duca degli Abruzzi 24, 10129 Torino, Italy}

\author{G. Vedovato}
\affiliation{INFN, Sezione di Padova, Via Marzolo 8, 35131 Padova, Italy}

\author{A. Vinante}
\affiliation{INFN, Sezione di Padova, Via Marzolo 8, 35131 Padova, Italy}

\author{M. Bignotto}
\affiliation{Dipartimento di Fisica, Universit\`{a} di Padova, 35131 Padova, Italy}
\affiliation{INFN, Sezione di Padova, Via Marzolo 8, 35131 Padova, Italy}

\author{M. Cerdonio}
\affiliation{Dipartimento di Fisica, Universit\`{a} di Padova, 35131 Padova, Italy}
\affiliation{INFN, Sezione di Padova, Via Marzolo 8, 35131 Padova, Italy}

\author{P. Falferi}
\affiliation{Istituto di Fotonica e Nanotecnologie, CNR-Fondazione Bruno Kessler, 38100 Povo, Trento, Italy}
\affiliation{INFN, Gruppo Collegato di Trento, Sezione di Padova, 38100 Povo, Trento, Italy}

\author{N. Liguori}
\affiliation{Dipartimento di Fisica, Universit\`{a} di Trento, 38100 Povo, Trento, Italy}
\affiliation{INFN, Gruppo Collegato di Trento, Sezione di Padova, 38100 Povo, Trento, Italy}

\author{S. Longo}
\affiliation{Dipartimento di Fisica, Universit\`{a} di Padova, 35131 Padova, Italy}
\affiliation{INFN, Sezione di Padova, Via Marzolo 8, 35131 Padova, Italy}

\author{R. Mezzena}
\affiliation{Dipartimento di Fisica, Universit\`{a} di Trento, 38100 Povo, Trento, Italy}
\affiliation{INFN, Gruppo Collegato di Trento, Sezione di Padova, 38100 Povo, Trento, Italy}

\author{A. Ortolan}
\affiliation{INFN, Laboratori Nazionali di Legnaro, 35020 Legnaro, Padova, Italy}

\author{G.A. Prodi}
\affiliation{Dipartimento di Fisica, Universit\`{a} di Trento, 38100 Povo, Trento, Italy}
\affiliation{INFN, Gruppo Collegato di Trento, Sezione di Padova, 38100 Povo, Trento, Italy}

\author{F. Salemi}
\affiliation{Dipartimento di Fisica, Universit\`{a} di Trento, 38100 Povo, Trento, Italy}
\affiliation{INFN, Gruppo Collegato di Trento, Sezione di Padova, 38100 Povo, Trento, Italy}

\author{L. Taffarello}
\affiliation{INFN, Sezione di Padova, Via Marzolo 8, 35131 Padova, Italy}

\author{S. Vitale}
\affiliation{Dipartimento di Fisica, Universit\`{a} di Trento, 38100 Povo, Trento, Italy}
\affiliation{INFN, Gruppo Collegato di Trento, Sezione di Padova, 38100 Povo, Trento, Italy}

\author{J.-P. Zendri }
\affiliation{INFN, Sezione di Padova, Via Marzolo 8, 35131 Padova, Italy}

\begin{abstract}
We analyze heat and work fluctuations in the gravitational wave detector AURIGA, modeled as a macroscopic electromechanical oscillator in contact with a thermostat and cooled by an active feedback system. The oscillator is driven to a steady state by the feedback cooling, equivalent to a viscous force. The experimentally measured fluctuations are in  agreement with our theoretical analysis based on a stochastically driven Langevin system. The asymmetry of the fluctuations of the absorbed heat characterizes the oscillator's nonequilibrium steady state and reveals the extent to which a feedback cooled system departs from equilibrium in a statistical mechanics perspective.
\end{abstract}
\maketitle

Cold damping feedback efficiently reduces the thermal noise motion of an oscillator by applying a viscous force. Since its first application in electrometers \cite{milatz}, it succeeded in a wide variety of devices \cite{karrai,schwab}, from nano to macroscopic resonators, and in a variety of implementations, including both optical and electrical forces. In basic research, the cold damping is considered in order to reduce below the level of intrinsic quantum fluctuations the uncertainty due to thermal noise of the position of macroscopic bodies \cite{aspelmeyer}, and to improve the behavior of gravitational wave detectors \cite{mancini}.
In this work we experimentally investigate the fluctuations of thermodynamic quantities of a cold-damped electromechanical oscillator: the resonant-bar gravitational wave detector AURIGA \cite{auriga}. In particular we verify that they are consistent with recent theories of nonequilibrium phenomena.

After the seminal works of Ref. \cite{fluctuations}, which introduced the Fluctuation Relation (FR) concerning the Probability Density Function (PDF) of the entropy production rate in nonequilibrium systems, a large number of papers has been devoted to similar problems (see for instance Ref. \cite{devel} for a review). One finds that the FR for some properly identified observable (called dissipation function) is quite generally valid in systems of physical interest \cite{rondoni}. After the experimental evidence obtained for dragged colloidal particles \cite{wang02}, electrical circuits \cite{garnier05} and mechanical oscillators \cite{douarche06}, the FR has become a standard tool to characterize nonequilibrium systems.
Here, following Ref. \cite{powerinjection} we focus on the FR for the PDF of the power necessary to maintain a dissipative system in a nonequilibrium steady state (NESS). A specific FR, obtained for the fluctuations of the injected power in a stochastically driven Langevin system \cite{farago02}, was recently confirmed in a simple electrical realization of that model \cite{falcon09}. Actually, this FR accurately fits the fluctuations of the injected power in
wave turbulence as well \cite{falcon}. We show that also the AURIGA detector, which is maintained in a NESS by an external driving in a feedback cooling scheme, can be described as a mechanical oscillator forced by a stochastic driving. We then analyze its behavior and demonstrate that: a) the statistics of its thermodynamic variables show a characteristic asymmetry between positive and negative fluctuations; b) the statistics of the injected power are in agreement with the FR of Ref. \cite{farago02}. These results 
reveal the extent to which a feedback cooled system departs from equilibrium in a statistical mechanics perspective and prove the limits of usual assumption that cold-damped oscillators at temperature $T_0$ are equivalent to higher-loss ones, in thermodynamic equilibrium at a temperature $T_{\textrm{eff}}<T_0$. 

\begin{figure}[t!]
\includegraphics[width=8.6cm,height=35mm]{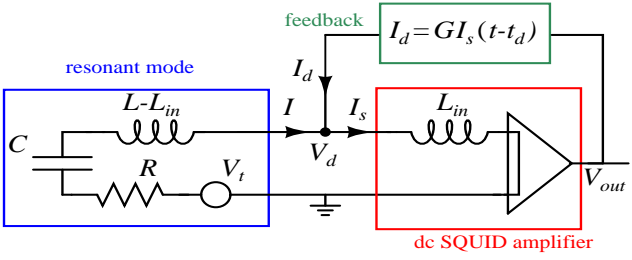}
\caption{(Color online). The normal mode is approximated, around its resonance frequency, by a series-RLC circuit. The dc SQUID is represented as current amplifier. The observable is the current $I_s$, and the electronic feedback cooling is obtained by sending back a current $I_d$ which is a delayed copy of $I_s$ reduced by $G\ll 1$. The SQUID output voltage is $V_{out}=A I_s$ with $A=2.6\,10^6\;\Omega$.
\label{fig:schema}}
\end{figure}

AURIGA is based on a $2.2\times 10^3$ kg, 3 m long bar made of a low-loss aluminum alloy (Al5056), cooled to liquid helium temperature $T_0=(4.6\pm 0.2)$\;K. The fundamental longitudinal mode of the bar, sensitive to gravitational waves, has effective mass M=$1.1 \times 10^3$ kg and resonance frequency $\omega_0/2\pi \sim 900$ Hz. The bar resonator motion is detected by a capacitive transducer followed by a double stage dc-SQUID amplifier; the displacement sensitivity is about $5 \times 10^{-20}$ m/$\sqrt{\textnormal{Hz}}$ over a $\sim$ 100 Hz bandwidth around $\omega_0$, largely limited by thermal noise. The detector can be
modeled by three coupled low-loss resonators: two mechanical ones (the bar and a plate of the capacitive transducer) and an LC electrical one \cite{baggio,tricarico}. Their dynamics is described by three normal modes at separate frequencies, 865.7, 914, 953 Hz, with quality factors respectively of $1.2\times 10^6,\;0.88\times 10^6,\;0.77 \times 10^6$, determined by mechanical losses in the bar and the transducer and by dielectric losses in the electrical components. Each mode is modeled as a RLC series electrical oscillator with an effective inductance $L$, capacitance $C$ and resistance $R$, which assume different values for the 3 oscillators (Fig.\,\ref{fig:schema}). These modes are electromechanical rather than purely mechanical, but each one collects a significant fraction of the energy of the two mechanical resonators. From the detector calibration we estimate that the energy injected by an impulsive excitation of the longitudinal mode of the bar is shared by the 3 modes in the ratio 48:36:16.

To the sole purpose of improving the electronics stability and easing the data analysis, AURIGA employs an electronic feedback cooling scheme on the detector (Fig.~\ref{fig:schema}), which is equivalent to a viscous force on the oscillators~\cite{vinante}. The dynamics of each electromechanical oscillator is described by the equations:
\begin{subequations}
\label{eq:prima}
\begin{eqnarray}
\label{eq:langevin}
&& \!\!\!\!\!\!\!\!\!\!\!\!\! (L-L_{in})\frac{d^2q(t)}{dt^2}+ R \frac{dq(t)}{dt}+ \frac{q(t)}{C} =V_T(t)- V_d(t)\\
\label{eq:Vd}&&\!\!\!\!\!\!\!\!\!\!\!\!\! V _d(t)= L_{in}\frac{dI_s(t)}{dt}\\
\label{eq:correnti}&&\!\!\!\!\!\!\!\!\!\!\!\!\! I(t)+I_d(t)= I_s(t)
\end{eqnarray}
\end{subequations}
where the current $I_s$ is the observable, $q$ is the charge on the capacitor, $I=\frac{dq(t)}{dt}$ the current through the inductance $L$, $V_d$ is the voltage at the node where the feedback is applied and $L_{in}$ is the input inductance of the SQUID amplifier. In thermodynamic equilibrium, each oscillator is driven by the stochastic voltage:
$V_T(t)= \sqrt{2k_BT_0 R}\; \Gamma(t)\,$,
where $\Gamma$ is a Gaussian white process. This should hold even in our nonequilibrium case, since the feedback cooling concerns only 3 modes, out of the very many degrees of freedom of the thermal bath, and is not expected to significantly affect the thermal noise due to the bath.   The noise due to the SQUID can be neglected at moderate feedback gains,  as those used in this experiment.
The very high quality factor of the oscillators imply that the currents $I_s$, $I_d$ and $I$ oscillate at $\omega_r=1/\sqrt{LC}$ with amplitude and phase changing appreciably only on timescales of several cycles. Thus, the quasi-harmonic approximation $I(t)=\hat{I}(t) \sin [\omega_r t+\hat{\phi}(t)]$, like the analogous ones for $I_s$ and $I_d$, seems appropriate. Operatively this is implemented by considering the signal only in a narrow frequency band around $\omega_r$.
Near resonance, a feedback force equivalent to a viscous damping can be obtained with the feedback current:
\begin{equation}
\label{eq:Idfeed}
I_d(t)=G I_s(t-t_d)
\end{equation}
where $t_d = \frac{\pi}{2 \omega_r}$ and $G\ll 1$. 
Equation (\ref{eq:Vd}) includes memory effects due to contributions from times $t-t_d$, because of the constraints (\ref{eq:correnti}) and (\ref{eq:Idfeed}); in the quasi-harmonic  approximation we have $I_s(t-t_d)\simeq \omega_r q_s(t)$. Hence each oscillator obeys:
\begin{equation}
\label{eq:langevinsystem}
 L\frac{dI_s(t)}{dt}+ I_s(t)\left[R + R_d \right]+
\frac{q_s(t)}{C} =\sqrt{2k_BT_0 R}\,\Gamma(t) 
\end{equation}
with $I_s(t)=\frac{dq_s(t)}{dt}$. Here $R_d=G \omega_r L_{in}$ expresses the viscous damping on the oscillator due to the feedback loop; the feedback efficiency is defined as $g=R_d/R$. 
The quasi-harmonic approximation is valid as long as the feedback damped oscillator is still a low loss one. In Eq. \eqref{eq:langevinsystem} the driving is the same white process of Eq. \eqref{eq:prima}: this is confirmed experimentally  by the Lorentz-shaped power spectrum of the current $I_s$ around the resonance \cite{vinante}.
Equation \eqref{eq:langevinsystem} is not invariant under time reversal ($q'_s=q_s, \,I'_s=-I_s$, $t'=-t$) and does not satisfy the Einstein relation. Nevertheless, it is \emph{formally} identical to that describing an oscillator with damping $R+R_d$, in equilibrium at the fictitious ``effective temperature'' $T_{\textrm{eff}}=T_0/(1+g)$. The discrepancy between $T_{\textrm{eff}}$ and the thermal bath temperature $T_0$ reveals the nonequilibrium nature of the phenomenon. Hence, the feedback cooled oscillator is usually treated as an equilibrium system, with $T_{\textrm{eff}}$ derived from the experimental value of $\left\langle \hat{I}_s^2(t)\right\rangle= 2\frac{k_B T_{\textrm{eff}}}{L}$, even if no bath at $T_{\textrm{eff}}$ is present. 

\begin{figure}[t!]
\includegraphics[width=8.6cm,height=99.6mm]{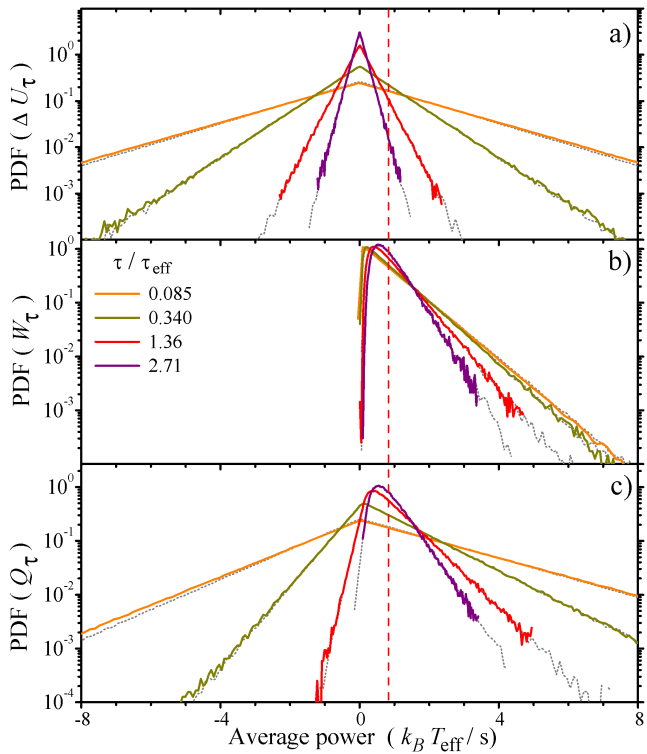}
\caption{(Color online). PDF (units of $s/(k_B T_{\textrm{eff}}$) of: (a) the time averaged energy difference $\Delta U_\tau$, (b) the work $W_\tau$ and (c) the heat $Q_\tau$ averaged at increasing values of the ratio $\tau/\tau_{\textrm{eff}}$. Data were collected by AURIGA in a 10 days time span. Dashed vertical line at 0.84$\;k_B T_{\textrm{eff}}$/s corresponds to the  mean value of $W_\tau$. The gray dotted lines are obtained by numerical simulation of Eq. \eqref{eq:langevinsystem} for a 50 days time span. The discrepancies observed at short $\tau$ between experimental and numerical data are within the uncertainty due to the experimental error in $\tau_{\textrm{eff}}$. 
 \label{fig:results}}
\end{figure}

Multiplying Eq.~\eqref{eq:langevinsystem} by $I_s(t)$ and integrating between $t$ and $t+\tau$, in the quasi-harmonic approximation we get an expression for the average power $P_\tau=\frac{1}{\tau} \int_{t}^{t+\tau} I_s(t') V_T(t') dt'$ injected by the stochastic thermal force during a time $\tau$:
\begin{equation}
\label{eq:power}
P_\tau =\Delta U_\tau + \frac{R+R_d}{\tau} \int_{t}^{t+\tau} I^2_s(t')dt'
\end{equation}
where $\Delta U_\tau=\frac{U(t+\tau)-U(t)}{\tau}$, $U(t)$ being the stored energy:
\begin{equation}
\label{eq:u0}
U(t)= \frac{1}{2}L I_s^2(t) + \frac{1}{2} \frac{q_s^2(t)}{C} =\frac{1}{2}L \hat{I}_s^2(t)
\end{equation} 
The term proportional to $R$ represents the heat dissipated by the oscillator toward the bath
while that proportional to $R_d$ is the work done by the oscillator on the feedback:
\begin{equation}
\label{eq:lavoroappr}
W_\tau= -\frac{1}{\tau}\int_{t}^{t+\tau} I_d(t') V_d(t') dt' = \frac{R_d}{\tau} \int_{t}^{t+\tau} I^2_s(t')dt'
\end{equation}
Notice that the last identity is strictly valid only within the quasi-harmonic approximation, which relates both $I_s(t-t_d)$ and $d{I}_s(t)/dt$ to the instantaneous current $I_s(t)$.
Further, if $\tau=N \frac{2\pi}{\omega_r}$, $N$ integer, in the same approximation we can also write:
\begin{equation}
\label{eq:lavoromediatointegrale}
W_\tau= \frac{1}{\tau}\frac{R_d}{2} \int_{t}^{t+\tau} \hat{I}^2_s(t')dt'
\end{equation}
By energy conservation we obtain the heat $ Q_\tau$, absorbed by the oscillator from the bath and averaged in a time $\tau$:
\begin{equation}
\label{eq:balance2}
Q_\tau=\Delta U_\tau + W_\tau
\end{equation}

To study nonequilibrium properties, we focused on the lowest frequency mode out of the 3, which is well separated in frequency from the other two and is thus our best approximation of a single oscillator.
The sampled current ${I}_s(t)$ was processed via the  AURIGA data analysis and integrated over the resonance in a 10 Hz bandwidth to obtain the current amplitude $\hat{I}_s(t)$ in the harmonic approximation. 
From dedicated calibration of AURIGA we measure $L=(1.67 \pm 0.01)\, 10^{-4}$\;H and $L_{in}=(1.48 \pm 0.01)\times 10^{-6}$\;H. A first set of data covers a continuous 10 days time span acquired in March 2008. They yield $\omega_r/2\pi =865.7\;$Hz, $T_{\textrm{eff}}= (21.1\pm 0.2)$\;mK and decay time $\tau_{\textrm{eff}}=(2.36 \pm 0.04)$\;s; hence we estimate $g=207\pm 10$, $R = (6.8\pm 0.5)\times 10^{-7}\;\Omega$ and $G=(1.74\pm 0.06)\times 10^{-2}$. The quality factor $\omega_r \tau_{\textrm{eff}}/2\simeq 6.5\times 10^3$ is high enough to justify the quasi-harmonic approximation leading to Eq. \eqref{eq:langevinsystem}.
Figure \ref{fig:results}a and \ref{fig:results}b show the PDF of the energy difference $\Delta U_\tau$ and of the work done by the oscillator $W_\tau$ averaged over growing times $\tau$: they are calculated from Eqs. \eqref{eq:u0} and \eqref{eq:lavoromediatointegrale} after dividing the experimental data in contiguous time intervals of duration $\tau$. Figure \ref{fig:results}c shows the corresponding heat $Q_\tau$ exchanged by the oscillator with the bath averaged over the time $\tau$, computed via the energy conservation Eq. \eqref{eq:balance2}.  The fluctuations of $Q_\tau$ are asymmetric, as expected for a NESS. Figure \ref{fig:results} shows also  excellent agreement with numerical simulations of Eq. \eqref{eq:langevinsystem}.

\begin{figure}[b!]
\includegraphics[width=8.6cm,height=62.5mm]{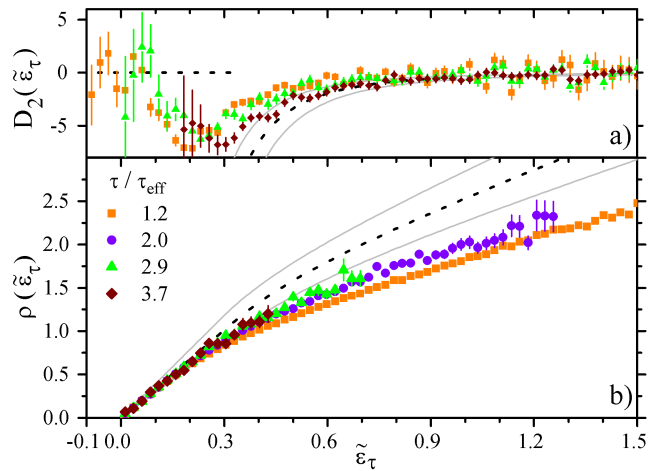}
\caption{(Color online). Comparison between theory (dashed lines) and experimental data at increasing values of the ratio  $\tau/\tau_{\textrm{eff}}=1.2,\;2.0,\;2.9,\;3.7$: the dataset are respectively 828, 768, 729, 457 days long. 
a) Plot of second derivative $D_2(\tilde{\epsilon}_\tau)$; the experimental data of $D_2$ for $\tau/\tau_{\textrm{eff}}= 2.0$ are not shown for clarity.
b) Plot of  $\rho(\tilde{\epsilon}_\tau)$.
Vertical error bars on experimental points of both plots come from statistical uncertainty. The shaded areas on the theoretical curves represent the uncertainty due to the experimental errors on the parameters, $\tau_{\textrm{eff}}$, $T_{\textrm{eff}}$ and $T_0$.
\label{fig:farago}}
\end{figure}

The PDF of $\Delta U_\tau$ is symmetric with respect to zero as for an equilibrium oscillator. It has exponential tails which decay faster for longer $\tau$. The PDF of $W_\tau$ is highly asymmetric. From Eq. \eqref{eq:lavoromediatointegrale} and $ \langle \hat{I}_s^2(t) \rangle= 2{k_B T_{\textrm{eff}}}/{L}$ we infer that $W_\tau$ is positive and has mean value $ \simeq 0.84\; k_B T_{\textrm{eff}}$/s. Hence, $Q_\tau$ takes negative values only for short integration times, with the characteristic time scale given by the cold damped oscillator decay time $\tau_{\textrm{eff}}=2 L/(R+R_d)$. For $\tau \gg \tau_{\textrm{eff}}$ the contribution of the time averaged energy is negligible. So in the presence of feedback ($R_d > 0$) there is a net heat transfer from bath to oscillator: this is the energy flux that feeds the NESS and makes the PDF of the heat asymmetric. On the contrary, if the feedback were switched off, we would have $R_d=0$ and $W_\tau =0$, hence $Q_\tau=\Delta U_\tau$ by Eq. \eqref{eq:balance2}. In this case the PDF of $Q_\tau$ would be symmetric with respect to its (zero) mean value, as in Fig. \ref{fig:results}a, but with $T_{\textrm{eff}}=T_0$.

The PDF of the injected power $P_\tau$ is essentially identical to that of $Q_\tau$, shown in Fig. \ref{fig:results}c, since $P_\tau\approx Q_\tau $ when  $g \gg 1$.  Notice that of the two terms in Eq. \eqref{eq:power}, only $\Delta U_\tau$ is responsible for the negative values of $P_\tau$. Thus, large positive values of $P_\tau$ are dominated by the contribution of the dissipated power [the integral in \eqref{eq:power}] more than they are for small values of $P_\tau$. The transition between these two regimes affects the shape of the PDF, which  has been calculated in Ref. \cite{farago02}. In a limit of large integration times it obeys: 
\begin{equation}\label{eq:flarge}
f(\tilde{\epsilon}_\tau) \equiv  \lim_{\tau \rightarrow \infty} \frac{\ln \textnormal{PDF}(\tilde{\epsilon}_\tau)}{\tau}
= \begin{cases}
-\gamma(1-2 \tilde{\epsilon}_\tau) 
&\text{if $\tilde{\epsilon}_\tau \le \frac{1}{3}$}\\
-\frac{\gamma}{4 \tilde{\epsilon}_\tau} (\tilde{\epsilon}_\tau -1)^2  
&\text{if $\tilde{\epsilon}_\tau\ge \frac{1}{3}$}
\end{cases}
\end{equation}
where $\tilde{\epsilon}_\tau=P_\tau L/(k_B T_0 R)$ is the reduced injected power and $\gamma=(R+R_d)/L=2/\tau_{\textrm{eff}}$.
A remarkable singularity, located at $\tilde{\epsilon}_\tau=1/3$, is present in the second derivative of $f(\tilde{\epsilon}_\tau)$. In Fig. \ref{fig:farago}a we plot the quantity $D_2(\tilde{\epsilon}_\tau)\equiv \frac{\partial^2 f(\tilde{\epsilon}_\tau)}{\partial \tilde{\epsilon}_\tau^2}$ evaluated from the output of AURIGA in the timespan May 2005/May 2008; here $T_{\textrm{eff}}= (22\pm 1)$\;mK and $\tau_{\textrm{eff}}=(2.4 \pm 0.2)$\;s. A valley is clearly visible in the experimental data, which we interpret as precursor of the asymptotic singularity.  The agreement with the asymptotic theory consistently improves as $\tau/\tau_{\textrm{eff}}$ grows. This indicates that the asymptotic relation Eq. \eqref{eq:flarge} holds even in presence of a harmonic pinning potential \cite{farago02}.

Eq. \eqref{eq:flarge} easily leads to the FR for the injected power, i.e. to the ratio between the probability of positive and negative fluctuations of $\tilde{\epsilon}_\tau$. If  we define $\rho(\tilde{\epsilon}_\tau) =^{\textrm{\;lim}}_{\tau \rightarrow \infty} \frac{1}{\tau} \ln \frac{\textnormal{PDF} (\tilde{\epsilon}_\tau)}{\textnormal{PDF} (-\tilde{\epsilon}_\tau)}$, we have:
\begin{equation}\label{eq:faragoratio}
\rho(\tilde{\epsilon}_\tau)  = \begin{cases}
4\gamma \tilde{\epsilon}_\tau, &\text{if $\tilde{\epsilon}_\tau<\frac{1}{3}$;}\\
\gamma \tilde{\epsilon}_\tau \left(\frac{7}{4}+ \frac{3}{2\tilde{\epsilon}_\tau}-\frac{1}{4\tilde{\epsilon}_\tau^2}\right), &\text{if $\tilde{\epsilon}_\tau\ge \frac{1}{3}$.}
\end{cases}
\end{equation}
As shown in Fig. \ref{fig:farago}b, positive values of $\tilde{\epsilon}_\tau$ are exponentially more probable than negative ones. Two conflicting features  determine the details of the experimental curves: the agreement with the asymptotic theory improves with $\tau/\tau_{\textrm{eff}}$, but the statistics blur at large values of $\tilde{\epsilon}_\tau$ since negative events are rarer. For this reason a slope change is clearly seen for small values of $\tau/\tau_{\textrm{eff}}$ where the precursor of the singularity occurs as shown in Fig. \ref{fig:farago}a, while it becomes barely visible at longer integration times.

In conclusion, we demonstrate that the actively cooled AURIGA detector is well described by the Langevin model of Eq.~\eqref{eq:langevinsystem}, which led us to evaluate the power $P_\tau$ injected by the stochastic thermal force, the work $W_\tau$ done on the feedback and the heat   $Q_\tau$ exchanged with the thermal bath. The statistics of $P_\tau$ are consistent with  Eq.~\eqref{eq:flarge}, and with the consequent nonlinearity of the FR. The fluctuations of $Q_\tau$ are asymmetric as expected for a NESS.

We acknowledge the contribution of the European Research Council within the 7th Framework Programme (FP7) of the European Community (EC), ERC Grant Agreement n. 202680. The EC is not responsible for any use that might be made of the data appearing therein.

\end{document}